\documentclass[epj]{svjour}

\usepackage{graphicx}
\usepackage{fancyhdr}

\def\lsim{\mathrel{\raise.3ex\hbox{$<$\kern-.75em\lower1ex\hbox{$\sim$}}}}

\def\gsim{\mathrel{\raise.3ex\hbox{$>$\kern-.75em\lower1ex\hbox{$\sim$}}}}

\newcommand{\nc}{\newcommand}
\nc{\renc}{\renewcommand}

\nc{\advp}[3]{{\it  Adv.\ in\ Phys.\ }{{\bf #1} {(#2)} {#3}}}
\nc{\annp}[3]{{\it  Ann.\ Phys.\ (N.Y.)\ }{{\bf #1} {(#2)} {#3}}}
\nc{\apl}[3]{{\it  Appl. Phys. Lett. }{{\bf #1} {(#2)} {#3}}}
\nc{\apj}[3]{{\it  Ap.\ J.\ }{{\bf #1} {(#2)} {#3}}}
\nc{\apjl}[3]{{\it  Ap.\ J.\ Lett.\ }{{\bf #1} {(#2)} {#3}}}
\nc{\app}[3]{{\it Astropart.\ Phys.\ }{{\bf #1} {(#2)} {#3}}}
\nc{\cmp}[3]{{\it  Comm.\ Math.\ Phys.\ }{{ \bf #1} {(#2)} {#3}}}
\nc{\cqg}[3]{{\it  Class.\ Quant.\ Grav.\ }{{\bf #1} {(#2)} {#3}}}
\nc{\epl}[3]{{\it  Europhys.\ Lett.\ }{{\bf #1} {(#2)} {#3}}}
\nc{\ijmp}[3]{{\it Int.\ J.\ Mod.\ Phys.\ }{{\bf #1} {(#2)} {#3}}}
\nc{\ijtp}[3]{{\it Int.\ J.\ Theor.\ Phys.\ }{{\bf #1} {(#2)} {#3}}}
\nc{\jmp}[3]{{\it  J.\ Math.\ Phys.\ }{{ \bf #1} {(#2)} {#3}}}
\nc{\jpa}[3]{{\it  J.\ Phys.\ A\ }{{\bf #1} {(#2)} {#3}}}
\nc{\jpc}[3]{{\it  J.\ Phys.\ C\ }{{\bf #1} {(#2)} {#3}}}
\nc{\jap}[3]{{\it J.\ Appl.\ Phys.\ }{{\bf #1} {(#2)} {#3}}}
\nc{\jpsj}[3]{{\it J.\ Phys.\ Soc.\ Japan\ }{{\bf #1} {(#2)} {#3}}}
\nc{\lmp}[3]{{\it Lett.\ Math.\ Phys.\ }{{\bf #1} {(#2)} {#3}}}
\nc{\mpl}[3]{{\it  Mod.\ Phys.\ Lett.\ }{{\bf #1} {(#2)} {#3}}}
\nc{\ncim}[3]{{\it  Nuov.\ Cim.\ }{{\bf #1} {(#2)} {#3}}}
\nc{\np}[3]{{\it  Nucl.\ Phys.\ }{{\bf #1} {(#2)} {#3}}}
\nc{\npps}[3]{{\it  Nucl.\ Phys.\ Proc.\ Suppl.\ }{{\bf #1} {(#2)} {#3}}}
\nc{\pr}[3]{{\it Phys.\ Rev.\ }{{\bf #1} {(#2)} {#3}}}
\nc{\pra}[3]{{\it  Phys.\ Rev.\ A\ }{{\bf #1} {(#2)} {#3}}}
\nc{\prb}[3]{{\it  Phys.\ Rev.\ B\ }{{{\bf #1} {(#2)} {#3}}}}
\nc{\prc}[3]{{\it  Phys.\ Rev.\ C\ }{{\bf #1} {(#2)} {#3}}}
\nc{\prd}[3]{{\it  Phys.\ Rev.\ D\ }{{\bf #1} {(#2)} {#3}}}
\nc{\prl}[3]{{\it Phys.\ Rev.\ Lett.\ }{{\bf #1} {(#2)} {#3}}}
\nc{\pl}[3]{{\it  Phys.\ Lett.\ }{{\bf #1} {(#2)} {#3}}}
\nc{\prep}[3]{{\it Phys.\ Rep.\ }{{\bf #1} {(#2)} {#3}}}
\nc{\prsl}[3]{{\it Proc.\ R.\ Soc.\ London\ }{{\bf #1} {(#2)} {#3}}}
\nc{\ptp}[3]{{\it  Prog.\ Theor.\ Phys.\ }{{\bf #1} {(#2)} {#3}}}
\nc{\ptps}[3]{{\it  Prog\ Theor.\ Phys.\ suppl.\ }{{\bf #1} {(#2)} {#3}}}
\nc{\physa}[3]{{\it  Physica\ A\ }{{\bf #1} {(#2)} {#3}}}
\nc{\physb}[3]{{\it  Physica\ B\ }{{\bf #1} {(#2)} {#3}}}
\nc{\phys}[3]{{\it Physica\ }{{\bf #1} {(#2)} {#3}}}
\nc{\rmp}[3]{{\it  Rev.\ Mod.\ Phys.\ }{{\bf #1} {(#2)} {#3}}}
\nc{\rpp}[3]{{\it Rep.\ Prog.\ Phys.\ }{{\bf #1} {(#2)} {#3}}}
\nc{\sjnp}[3]{{\it Sov.\ J.\ Nucl.\ Phys.\ }{{\bf #1} {(#2)} {#3}}}
\nc{\spjetp}[3]{{\it Sov.\ Phys.\ JETP\ }{{\bf #1} {(#2)} {#3}}}
\nc{\yf}[3]{{\it Yad.\ Fiz.\ }{{\bf #1} {(#2)} {#3}}}
\nc{\zetp}[3]{{\it Zh.\ Eksp.\ Teor.\ Fiz.\  }{{\bf #1}  {(#2)} {#3}}}
\nc{\zp}[3]{{\it Z.\ Phys.\ }{{\bf #1} {(#2)} {#3}}}
\nc{\ibid}[3]{{\sl ibid.\ }{{\bf #1} {#2} {#3}}}
\def\mytitle{My title} 
\def\myauthors{My name}  
\def\mytype{My type of session}
\def\mysession{My session}

\def\mytitle{RH Sneutrino Condensate CDM and the Baryon-to-Dark Matter Ratio} 
\def\myauthors{John McDonald}    
\def\mytype{Contributed Talk}    
\def\mysession{Cosmology and Astrophysics}

\begin{document}
\title{RH Sneutrino Condensate CDM and the Baryon-to-Dark Matter Ratio}
\subtitle{}
\author{John McDonald\inst{1}
}                     
\institute{Cosmology and Astroparticle Physics Group, Dept. of Physics, Lancaster University, UK}
%
\date{}
\abstract{ The similarity of the observed mass densities of 
baryons and cold dark matter may be a sign they have a related origin. The baryon-to-dark matter ratio can be understood in the MSSM with right-handed (RH) neutrinos if CDM is due to a d = 4 flat direction condensate of very weakly coupled RH sneutrino LSPs and the baryon asymmetry is generated by Affleck-Dine leptogenesis along a d = 4 $\left(H_{u}L\right)^2$ flat direction. Observable signatures of the model include CDM and baryon isocurvature perturbations and a distinctive long-lived NLSP phenomenology. 
\PACS{
      {98.80.Cq}{Cosmology}  
     } 
} 
\maketitle
\section{Introduction}
\label{intro}

      A striking feature of the observed Universe is the similar mass density in baryons and cold dark matter. (The 'Baryon-to-Dark Matter' (BDM) ratio.) From the WMAP three-year results for the $\Lambda$CDM model, 
$\Omega_{DM}/\Omega_{B} = 5.65 \pm 0.58$ \cite{wmap}. However, in most models the physics of baryogenesis and dark matter 
production are unrelated. So why is the mass density in baryons within an order of magnitude of that of dark matter?  
There are three possibilities:
\newline (i)  A remarkable coincidence.
\newline (ii) Some anthropic selection mechanism, usually assumed but undefined (e.g. in the case of thermal relic neutralino dark matter).
\newline (iii) The mechanisms for the origin of the baryon asymmetry and dark matter are related. 

   The latter possibility seems the simplest interpretation of the BDM ratio. Indeed, we may be ignoring a {\it big clue} to the nature of the correct BSM particle theory. It is highly non-trivial for a particle physics theory to have within its structure (without contrivance) a mechanism that can account for the BDM ratio. Therefore if the BDM ratio is due to such a mechanism it would provide us with a powerful principle by which to select the best canadiate
particle physics models.

             BDM models broadly divide into two classes:
\newline {\bf 1).Charge conservation based:} The dark matter particle and baryon number are related by a conserved charge, $Q_{B} + Q_{cdm} = 0 \Rightarrow n_{cdm} \sim n_{B}$. The CDM particle mass satisfies 
$ m_{cdm} = m_{n}n_{B}/n_{cdm}$
and so $m_{cdm} \sim 1 GeV$ is necessary. However, this does fit well with SUSY if the LSP mass is $O(m_{W})$ or larger. 
\newline {\bf 2). Dynamics based:} In this case the dark matter and baryon densities are related by {\it similar} physical mechanisms for their origin. This implies a less rigid relation between $n_{B}$ and $n_{CDM}$, which may allow us to understand why it is the mass rather than number densities that are observed to be similar.

\section{SUSY BDM Models}
\label{sec:2}

        It is perhaps significant that many of the BDM models studied in the past are based on SUSY and the Affleck-Dine (AD) baryogenesis mechanism \cite{susybdm}. Moreover, the SUSY BDM models tend to be far less baroque than the non-SUSY schemes \cite{nsbdm}. So SUSY already appears favoured by the 'BDM principle'. However, the previously considered schemes have all been of the charge conservation type, which are not compatible with SUSY breaking schemes which have $O(m_{W})$ soft SUSY-breaking masses. 
So is there a compelling  SUSY dynamical BDM model? In \cite{dcdm} we suggested that there is: 
{\bf \begin{center} \underline{Dynamical SUSY BDM Model:} \end{center} 
\begin{center} d=4 $\left(H_{u}L\right)^2$ Affleck-Dine leptogenesis \end{center} 
\begin{center} +  \end{center} 
\begin{center} RH Sneutrino Condensate CDM with d=4 superpotential\end{center}  }
\noindent The key ingredient is that the potentials of the AD scalar and the RH sneutrino condensate are both 
lifted by non-renormalizable terms of the same mass dimension, in this case $d = 4$. In this case the 
number densities in baryons and dark matter are related, but not by the tight condition of charge conservation \cite{dcdm}. This allows 
the dynamics to account for $n_{B} \sim 100 n_{cdm}$ and so why the mass densities are related. 

\begin{figure}
\centering 
    \includegraphics[width=2.5in, angle = -0]{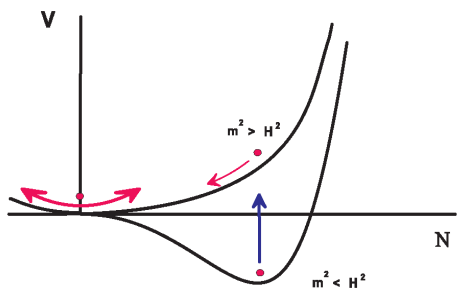} 
  \caption{RH Sneutrino oscillations} 
  \label{fig2} 
\end{figure}

The RH sneutrinos are introduced via the superpotential
$$  W = W_{MSSM} + W_{\nu}  \;\;,\;\;
 W_{\nu} = \lambda_{\nu} N H_{u}L + \frac{M_{N}}{2} N^2    ~.$$ 
If $M_{N} < M_{W}$ then the RH sneutrino can be the LSP\footnote{RH sneutrino dark matter with $M_{N} = 0$ was first considered in a thermal relic model in \cite{trsn}.}. We will consider the simplest case where 
$M_{N} = 0$ and the neutrinos are pure Dirac, with $\lambda_{\nu} < 10^{-13}$. The small coupling ensures that the 
RH sneutrinos are out of thermal equilibirium and so a condensate formed in the early Universe will exist today. 
The dynamics of the scalar fields follows the now-standard flat direction scenario (Figure \ref{fig2}). With a Planck-suppressed non-renormalizable term of the form ($M = M_{Pl}/\sqrt{8\pi}$)
$$ W = \frac{\lambda_{N} N^4}{4! M} ~,$$
the potential is (neglecting unimportant A-terms)  
$$ V(N) \approx \left( m_{N}^{2} - c_{N}H^{2} \right) |N|^{2} + \frac{\lambda_{N}^{2} |N|^{2\left(n-1\right)}}{M^{2\left(n-3\right)}}    ~,$$
where $m_{N} \approx 100 GeV$ is from SUSY breaking.  
This non-renormalizable term is of the lowest possible dimension and so perhaps the most natural one to consider. 
The RH sneutrino will begin to oscillate once $H^{2} = H^2_{osc} \approx m_{N}^{2}/c_{N}$, with initial amplitude 
$$ |N|_{osc} \approx   |N|_{min} [H \approx H_{osc \;N}] = \left(\frac{12}{\lambda_{N}^{2}}\right)^{1/4} \left(m_{N}M\right)^{1/2} 
~.$$
The resulting energy density today is then
$$\rho_{N\;o}  
= \frac{\sqrt{12} \pi^2}{45} \frac{c_{N}  T_{\gamma}^{3} 
T_{R} m_{N} }{\lambda_{N} M} ~,$$ 
and the observed dark matter density is obtained if the reheating temperature satisfies 
$$T_{R} \approx 2.6 \times 10^{7} \; \frac{\lambda_{N}}{c_{N}}
 \left(\frac{h}{0.7}\right)^{2} 
\left(\frac{\Omega_{N}}{0.23}\right) 
\left(\frac{100 GeV}{m_{N}}\right) GeV  ~.$$ 
This is consistent with the most recent thermal gravitino upper bound \cite{grav}, $T_{R} < 10^{6-8} GeV$, 
when \newline $\lambda_{N}/c_{N} \; ^{<}_{\sim} \; 0.1-1$.

        The Affleck-Dine mechanism is arguably the simplest and most natural way to generate the matter-antimatter 
asymmetry in the context of the MSSM.  The lowest dimension (B-L)-violating operator capable of doing this is 
$\left(H_{u}L\right)^{2}$. The corresponding $d=4$ flat direction potential is 
$$ V(\Phi) = \left( m_{\Phi}^{2} - c_{\Phi}H^{2}\right) 
\left|\Phi\right|^{2} $$ 
$$\;\;\;\;\;\;\;\; +  \left( A_{\Phi}\frac{ \lambda_{\Phi}}{4!M} \Phi^{4}
 + h.c. \right) + \frac{ \left|\lambda_{\Phi} 
\right|^{2}}{3!^{2} M^{2}} \left|\Phi\right|^{6}   ~.$$
At $cH^2 \sim m_{\Phi}^2$, the CP- and L-violating A-term kicks the 
real and imaginary oscillations out of phase, leaving the 
$\Phi$ field in a ellipitical orbit in the complex $\Phi$ plane. 
This corresponds to a L-asymmetry in the condensate, which is conserved 
once the $\Phi$ field and A-term diminishes via expansion. Once the scalars in the condensate decay perturbatively, spahleron processes will convert the L-asymmetry into an equivalent B-asymmetry, $n_{B} = -(8/23) n_{L\;initial}$. The resulting B-asymmetry is then \cite{dcdm}
$$   n_{B} \approx \frac{f_{A}}{4} \frac{\rho_{\Phi\;o}}{m_{\Phi}}\;\;;\;\; f_{A} = \frac{16}{23} \sin\left(2 \theta\right)
 \sin\left(\delta\right)    ~,$$
where $\theta$ and $\delta$ are CP-violating angles. 
The key point is that 
$$ \rho_{\Phi \; o} = m_{\Phi}^{2} \phi(t_{o})^{2}/2$$ 
is the density the $d=4$ $\Phi$ condensate would have in the Universe at present if the $\Phi$ condensate did not decay. {\it Therefore there is a direct connection between the 
B asymmetry generated along a $d=4$ flat direction and the dark matter 
density due to condensate lifted by a $d=4$ non-renormalizable term.} 
This connection is purely due to the similar dynamics of the scalar field-based mechanisms for the origins of the baryon asymmetry and of dark matter. As a result of its dynamical nature, the relationship between the baryon and dark matter number densities is looser than the strict relation found in charge conservation-based BDM models. The model has pluses and minuses. On the plus side, for reasonable values of the parameters $\lambda_{i},c_{i}$ we can account for the observed BDM ratio; a minus is that the parameter
dependence of the dynamics makes precise prediction impossible without a theory of $\lambda_{i},c_{i}$. However, these should eventually be calculable within a complete theory of Planck-scale physics.

  The resulting BDM ratio is given by 
$$\frac{\Omega_{B}}{\Omega_{DM}} \approx 
\frac{f_{A}}{4} \frac{m_{n}}{m_{\Phi}} 
\frac{\rho_{\Phi \; o}}{\rho_{N \; o}} 
 = \frac{f_{A}}{400} 
\left(\frac{100 GeV}{m_{N}}\right) 
\left[ \frac{c_{\Phi}}{c_{N}}  \frac{\lambda_{N}}{\lambda_{\Phi} }   \right] ~.$$ 
The square-bracketed terms reflect the dynamical nature of the BDM ratio 
in this model. If $c_{\Phi}/c_{N}$ and $\lambda_{N}/\lambda_{\Phi}$ are both $\sim 1$ then we would get the same kind of relation as in the charge-conservation type models, $n_{B}/n_{DM} \sim 1$. However, thanks to the freedom offered by the dynamics, the observed BDM ratio can be understood if there is a small hierarchy between 
$\lambda_{\Phi}$ and $\lambda_{N}$ e.g.
$$ \lambda_{\Phi} \sim 0.01 \lambda_{N}\;,\;\; f_{A} \sim 0.5\;,\;\; 
     c_{\Phi} \sim c_{N} \;,\;\; m_{N} \sim 100 GeV $$ 
$$ \Rightarrow  \frac{\Omega_{B}}{\Omega_{DM}}\approx \frac{1}{6}   ~.$$ 
It is worth emphasizing what has been gained here: 
\newline $\bullet$ The 'big mystery' of why completely unrelated mechanisms of baryogenesis and dark matter generation produce mass densities within an 
order of magnitude has been reduced to a simple hierarchy of couplings. Such a hierarchy of non-renormalizable Yukawas is entirely plausible, given the range of values observed for the Standard Model Yukawa couplings. 
\newline $\bullet$ No new physics has been added to the MSSM beyond the RH neutrinos required to account for neutrino masses. The Planck-suppressed non-renormalizable terms are of the lowest order and therefore what we would add   
to the model on general terms anyway. 

                  It is quite remarkable that the 
MSSM has within its structure such a simple and economical mechanism for generating and relating the dark matter and baryon densities. As such, the MSSM with neutrino masses appears to be strongly favoured by the 'BDM principle'. 
Note also the power of the BDM ratio as a selection rule: it not only identifies the particle physics model, but also the mechanisms of baryogenesis and dark matter generation and the identity of the dark matter particle itself.  

\section{Observable Features of the Model} 

\begin{figure}
\centering 
    \includegraphics[width=2.5in, angle = -0]{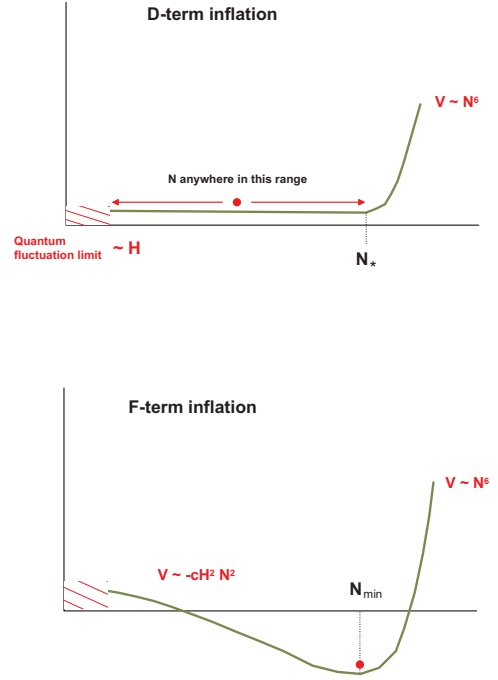} 
  \caption{RH Sneutrino potential during D- and F-term inflation. The $\Phi$ potential is similar. } 
  \label{fig1} 
\end{figure}

      The model has a number of potentially observable features:
\newline {\bf (i) CDM and Baryon Isocurvature perturbations.} These arise   either because the 
$N$ and $\Phi$ are effectively massless during inflation (as in D-term inflation models) or because the 
phase field is effectively massless (as in F-term inflation models with suppressed $H$ coreections to the A-terms)\cite{dcdm}. Therefore the observability of isocurvature perturbations depends on the nature of SUSY inflation. The CDM isocurvature (CDI) contribution to the CMB power spectrum is 
$$ C_{l} = (1 - \alpha) C_{l}^{ad} + \alpha C_{l}^{iso}$$ 
$$ \alpha_{CDI} \approx
\frac{H_{I}^{2}}{\pi^{2} P_{\cal{R}} N_{I}^{2}}$$ 
where $N_{I}$, $H_{I}$ are the values duing inflation. 
Using this, and comparing with the present upper bound \cite{bean}, 
$\alpha < 0.26$, we obtain an upper bound on 
$H_{I}$ 
$$H_{I} \;^{<}_{\sim} \; \left(\frac{48}{5}\right)^{1/2} \frac{\pi^{2}
 P_{{\cal R}} M \alpha_{lim}}{\lambda_{N}}  $$
$$ \equiv 4.4 \times 10^{11} 
 \left(\frac{0.1}{\lambda_{N}}\right) 
\left(\frac{\alpha_{lim}}{0.26}\right)
\left(\frac{N_{I}}{N_{*}}\right)^{2} GeV ~.$$
Here $N_{*}$ is the upper limit on $N_{I}$ where the non-renormalizable 
potential gives $N$ an effective mass of order $H_{I}^2$. (Figure \ref{fig1}.)  
The upper bound on $H_{I}$ is close to typical values in SUSY inflation models e.g. D-term hybrid inflation has $H_{I} \approx 1.1 \times 10^{13} g GeV$.

             In addition to CDM isocurvature perturbations of the RH sneutrino density, there can also be baryon isocurvature (BI) perturbations due to quantum fluctuations of $\Phi$. The ratio of baryon to CDM isocurvature perturbation is 
$$ \frac{\alpha_{BI}}{\alpha_{CDI}} 
= \frac{f_{\theta}^{2} N_{I}^{2}}{4 \phi_{I}^{2}} 
\left(\frac{\Omega_{B}}{\Omega_{CDM}} \right)^{2} $$ 
$$ \approx 
8 \times 10^{-3}  f_{\theta}^{2} \left( \frac{N_{I}}{\phi_{I}} \right)^{2} \;\;;\;\;\;\;  f_{\theta} = \frac{2}{\tan 2\theta} \sim 1  ~.$$ 
The CDM and baryon isocurvature perturbations cannot be distinguished 
by CMB observations, but could eventually be observed via the 21cm background \cite{21cm}. The ratio then becomes a direct indicator of the nature of SUSY inflation. F-term inflation implies $\Phi_{I}$ and $N_{I}$ are at the minimum of their potentials (Figure \ref{fig1}), so the baryon isocurvature contribution is negligible in this case \cite{dcdm} 
$$ \frac{\alpha_{BI}}{\alpha_{CDI}} 
 \approx 
8 \times 10^{-3}  f_{\theta}^{2} \left( \frac{c_{N}}{c_{\phi}} \right)^{1/2} \frac{\lambda_{\phi}}{\lambda_{N}} \sim 10^{-3}- 10^{-4}    ~.$$ 
On the other hand, in D-term inflation $N_{I}$ and $\Phi_{I}$ are undetermined, so a large or even dominant baryon isocurvature perturbation is possible in this case. 
\newline{\bf (ii) Long-lived NLSP collider pheomoenology}  The RH sneutrino 
can be regarded as a third member of the family of very weakly interacting SUSY dark matter candidates, joining the gravitino \cite{gravitino} and axino \cite{axino}. In common with these better-known candidates, MSSM collider pheomenology will be quite different from conventional (thermal-relic) neutralino LSPs, including the possibility of charged NLSPs. To distinguish RH sneutrino LSPs from gravitinos and axinos, the best possibility would be if the stau was the NLSP (MSSM-LSP) and if its decay mode to LSPs could be observed by trapping and observing its lifetime and final states. Final states for stau decay to gravitinos and axinos generally have $\tau^{-}$ leptons, whereas decay to RH sneutrinos typically will have a charged Higgs $h_{u}^{-}$. More detailed study of NLSP decays is called for, to ensure that the lifetime of the stau is short enough to avoid disrupting light element abundances. 
The NLSP phenomenology of RH sneutrino condensate CDM can also be distinguised from RH sneutrino dark matter from thermal relic NLSP decay \cite{trsn}, since the latter is restricted to the parameter region where $\Omega_{NLSP} > \Omega_{cdm}$, whereas there is no restriction on $\Omega_{NLSP}$ in the case of RH sneutrino condensate dark matter, allowing the MSSM-LSP to exist in the region of MSSM parameter space where $\Omega_{NLSP} \ll  \Omega_{cdm}$. Observation of a CDM isocurvature perturbation would also distinguish between condensate and thermal relic RH sneutrinos. 

\section{Summary} 

         RH sneutrino condensate CDM combined with Affleck-Dine baryogenesis can plausibly account for the observed similarity of the 
baryon and dark matter mass densities in the Universe. Seen as a selection principle, the requirement that a particle physics 
model can {\it without contrivance} account for the BDM ratio 
favours the MSSM with neutrino masses and  RH sneutrino condensate 
CDM. It is quite remarkable that the MSSM with neutrino masses has the ability to account for the BDM ratio as a natural consequence its structure. 

      CDM and baryon isocurvature perturbations are possible, the ratio of which gives information on the nature of SUSY inflation. Long-lived NLSP phenomenology is also expected, which can be 
distinguished from gravitino and axino LSP phenomenology via trapped stau final states, and from thermal relic RH sneutrino LSP phenomenology once the parameters of the MSSM are known.   

       There are a number of issues which remain to be addressed. One is the possibility that $M_{N} \neq 0$. In this case it is possible that the 
heavier generation condensate RH sneutrinos, which will have a lifetime longer than the age of the Universe so long as $\lambda_{\nu}$ is still small,  would decay into the LSP RH sneutrino plus $e^{+}e^{-}$. This could produce a potentially observable diffuse $\gamma$-ray background. In addition, the phenomenology and cosmology of MSSM-LSPs in this model should be studied in detail. 

\section*{Acknowledgement} This work was supported (in part) by the European Union through the Marie Curie Research and Training Network "UniverseNet" (MRTN-CT-2006-035863) and by STFC (PPARC) Grant PP/D000394/1.

%
%

\end{document}